\documentclass[aps,prl,reprint,superscriptaddress,showpacs,showkeys]{revtex4-1}
\usepackage[latin9]{inputenc}
\usepackage{babel}
\usepackage{amsmath}
\usepackage{amssymb}
\usepackage{graphicx}

\usepackage[title]{appendix}
\usepackage[unicode=true,
bookmarks=false,
 breaklinks=false,pdfborder={0 0 1},colorlinks=true,linkcolor=red,citecolor=blue]
 {hyperref}
\makeatletter
\@ifundefined{textcolor}{}
{%
 \definecolor{BLACK}{gray}{0}
 \definecolor{WHITE}{gray}{1} 
 \definecolor{RED}{rgb}{1,0,0}
 \definecolor{GREEN}{rgb}{0,1,0}
 \definecolor{BLUE}{rgb}{0,0,1}
 \definecolor{CYAN}{cmyk}{1,0,0,0}
 \definecolor{MAGENTA}{cmyk}{0,1,0,0}
 \definecolor{YELLOW}{cmyk}{0,0,1,0}
}

\usepackage{amsthm}
\usepackage{amsfonts}
\usepackage{bbm}
\usepackage{epsfig}
\usepackage{times}
\usepackage{babel}
\usepackage{color}
\usepackage{framed}
\usepackage{changes}
\usepackage{float}
\usepackage{array}
\makeatletter

\@ifundefined{textcolor}{}
{%
 \definecolor{BLACK}{gray}{0}
 \definecolor{WHITE}{gray}{1}
 \definecolor{RED}{rgb}{1,0,0}
 \definecolor{GREEN}{rgb}{0,1,0}
 \definecolor{BLUE}{rgb}{0,0,1}
 \definecolor{CYAN}{cmyk}{1,0,0,0}
 \definecolor{MAGENTA}{cmyk}{0,1,0,0}
 \definecolor{YELLOW}{cmyk}{0,0,1,0}
}

\usepackage{amsthm}
\usepackage{amsfonts}
\usepackage{bbm}
\usepackage{epsfig}
\usepackage{times}
\usepackage{babel}
\usepackage{color}
\usepackage{framed}
\usepackage{changes}
\usepackage{float}
\usepackage{array}

\makeatother

\begin{document}

\title{Critical scalarization and descalarization of black holes in a generalized scalar--tensor theory}

\author{Yunqi Liu}
\thanks{co-first authors}
\address{\textit{Center for Gravitation and Cosmology, College of Physical
Science and Technology, Yangzhou University, Yangzhou 225009, China}}

\author{Cheng-Yong Zhang}
\thanks{co-first authors}
\address{\textit{Department of Physics and Siyuan Laboratory, Jinan University,
Guangzhou 510632, China }}

\author{Qian Chen}
\address{\textit{School of Physical Sciences, University of Chinese Academy
of Sciences, Beijing 100049, China}}

\author{Zhoujian Cao}
\address{\textit{Department of Astronomy, Beijing Normal University, Beijing 100875, China }}
\address{\textit{School of Fundamental Physics and Mathematical Sciences,
Hangzhou Institute for Advanced Study, UCAS, Hangzhou 310024, China }}

\author{Yu Tian}
\address{\textit{School of Physical Sciences, University of Chinese Academy
of Sciences, Beijing 100049, China}}
\address{\textit{Institute of Theoretical Physics, Chinese Academy of Sciences,
Beijing 100190, China}}

\author{Bin Wang}
\thanks{corresponding author, wang\_b@sjtu.edu.cn}
\address{\textit{Center for Gravitation and Cosmology, College of Physical
Science and Technology, Yangzhou University, Yangzhou 225009, China}}
\address{\textit{Shanghai Frontier Science Center for Gravitational Wave Detection, Shanghai Jiao Tong
University, Shanghai 200240, China}}

\begin{abstract}
We study the critical dynamics in scalarization and descalarization in the fully nonlinear dynamical evolution in the class of theories with a scalar field coupling with both Gauss--Bonnet (GB) invariant and Ricci scalar.
We explore the manner in which the GB term triggers black hole (BH) scalarization.
A typical type I critical phenomenon is observed, in which an unstable critical solution emerges at the threshold and acts as an attractor in the dynamical scalarization.
For the descalarization, we reveal that a marginally stable attractor exists at the threshold of the first-order phase transition in shedding off BH hair.
This is a new type I critical phenomenon in the BH phase transition.
Implications of these findings are discussed from the perspective of thermodynamic properties and perturbations for static solutions.
We examine the effect of scalar--Ricci coupling on the hyperbolicity in the fully nonlinear evolution and observe that such coupling can suppress the elliptic region and enlarge parameter space in computations.
\end{abstract}


\maketitle

\section{Introduction}
One of the fascinating topics in general relativity is the black hole (BH) no-hair theorem. Recently, an attempt to challenge this theorem has been made by considering a quadratic scalar field coupling with the Gauss--Bonnet (GB) \citep{Doneva1711,Silva1711,Antoniou1711,Cunha:2019dwb,Herdeiro:2020wei,Berti:2020kgk,Dima:2020yac} or Maxwell invariant \citep{Herdeiro:2018wub}.
At the linear level, such coupling can trigger tachyonic instability, which spontaneously scalarizes a bald black hole (BBH) through a second-order phase transition.
Besides scalarization, dynamical descalarization through second-order phase transition was observed in BH binaries \cite{Silva:2020omi, Elley:2022ept}.
Considering higher order scalar coupling, in the fully nonlinear evolution, both scalarization and descalarization can occur more violently through a first-order phase transition in both the scalar--GB (sGB) \cite{Doneva:2021tvn,Blazquez-Salcedo:2022omw,Doneva:2022byd} and Einstein--Maxwell-scalar (EMS) theories \citep{Blazquez-Salcedo:2020nhs,LuisBlazquez-Salcedo:2020rqp,Blazquez-Salcedo:2020crd,Zhang:2021nnn,Zhang:2022cmu}.
Regarding the EMS theory, critical dynamics in different orders of phase transitions were thoroughly explored to realize scalarization and descalarization \citep{Zhang:2021nnn, Zhang:2022cmu}.
For an initial scalar wavepacket parameterized by $p$, which can represent the amplitude, wavepacket width, or length, there exists a threshold $p_*$.
When $p$ approaches $p_*$, a violent change occurs at very early stages, primarily driven by initial data.
Subsequently, all intermediate solutions are attracted to a critical solution (CS), which corresponds to a metastable, scalarized BH solution in theory, and remain in this state for an extended period.
By precisely fine-tuning $p$ to the critical value $p_*$, the evolution theoretically indefinitely remains on this CS.
In the late stages, the intermediate solutions decay to BBHs if $p<p_*$ or to scalarized charged BHs if $p>p_*$.
The critical phenomena observed in the first-order phase transition resemble those found in type I critical gravitational collapse \cite{Choptuik:1996yg,Liebling:1996dx,Bizon:1998kq,Gundlach:2007gc}.
The discovery of critical phenomena in scalarization and descalarization has supplemented the existing knowledge on critical dynamics in gravity.

It is intriguing to investigate the critical dynamics in a theory with coupling between a scalar field and the GB term and reveal the intrinsic mechanism of its scalarization and descalarization. A theory with scalar--GB coupling is of interest to astrophysics \cite{Shiralilou:2021mfl,Danchev:2021tew,Wong:2022wni,Maselli:2020zgv,Guo:2022euk,Barsanti:2022ana}; however, most astrophysical tests of scalarized BHs (SBHs) induced by the GB term are limited in stationary solutions and linear perturbations \cite{Blazquez-Salcedo:2018jnn,Silva:2018qhn,Hod:2020jjy}.
Numerically, the study of dynamics in a theory with scalar--GB coupling is more difficult than that in the EMS theory due to the lack of well-posedness of the Cauchy problem \cite{Ripley:2019hxt,Ripley:2019irj,Julie:2020vov,Witek:2020uzz,East:2020hgw}.
To avoid the ill-posed dynamical evolution, one limited the study in the weak coupling regime \cite{Doneva:2021tvn,Blazquez-Salcedo:2022omw,Witek:2018dmd,Silva:2020omi,Doneva:2021dqn,Doneva:2022byd,Elley:2022ept} or adopted a small parameter space to escape the problematic regions
\cite{Ripley:2020vpk, Kuan:2021lol, East:2021bqk, Liu:2022eri}.
A weak completion dubbed as fixing-the-equations was proposed to reduce the mathematical pathologies and allow one to push the evolution further \cite{Franchini:2022ukz}.
A perturbative analysis showed that scalar--Ricci (SR) coupling can reduce the region of parameter space where hyperbolicity breaks down \cite{Antoniou:2022agj}.
This coupling was proposed in a study on neutron star scalarization \citep{Damour1993,Damour1996,Harada1997}, and it was shown to be crucial for observational viability \cite{Antoniou:2020nax,Ventagli:2021ubn,Antoniou:2021zoy}.
It is worth examining the effect of such SR coupling on the hyperbolicity in the fully nonlinear evolution and studying its role in the critical dynamics in scalarization and descalarization in a theory with scalar--GB coupling.

In this letter, we study dynamical critical behaviors in scalarization and descalarization in a generalized theory with a scalar field coupling with both the Ricci scalar and GB invariant.
We show the manner in which the GB term triggers the nonlinear scalarization, transforms a linearly stable BBH into an unstable SBH attractor,
and finally transforms it into a stable SBH.
We demonstrate that the property of the CS in the scalarization qualitatively resembles that in the EMS theory.
Remarkably, we reveal novel critical behaviors in the dynamical descalarization.
The CS at the threshold of descalarization is marginally stable and acts as a marginal stable attractor, reflecting a new type I critical phenomenon in the BH phase transition.
We disclose the thermodynamic properties and perturbations for static solutions to help understand such novel phenomenon more effectively.
Furthermore, the effect of SR coupling on the hyperbolicity in the fully nonlinear evolution is investigated.

\section{Setup}
We consider a generalized scalar--tensor theory with action \citep{Antoniou:2020nax,Ventagli:2021ubn, Ventagli:2020rnx,Antoniou:2021zoy,Antoniou:2022agj}
\begin{equation}
S=\frac{1}{16\pi}\int d^{4}x\sqrt{-g}\left[R-\frac{1}{2}(\partial\phi)^{2}+f(\phi)\left(\beta R+\mathcal{G}\right)\right],\label{eq:action}
\end{equation}
where a scalar field $\phi$ non-minimally couples to both the Ricci scalar $R$ and GB invariant $\mathcal{G}=R^{2}-4R_{\mu\nu}R^{\mu\nu}+R_{\mu\nu\alpha\beta}R^{\mu\nu\alpha\beta}$ by $f(\phi)$; $\beta$ is a dimensionless parameter.
The BBH is a solution if $f(0)=0$, $\frac{df}{d\phi}(0)=0$.
The positive and sufficiently large $\frac{d^{2}f}{d\phi^{2}}(0) \mathcal{G}$ can trigger tachyonic instability and spontaneously scalarize the BBH.
Many studies observed that at the linear level, the coupling between a scalar field and the Ricci scalar can considerably influence BH scalarization \citep{ Ventagli:2020rnx,Antoniou:2021zoy,Antoniou:2022agj}.
We discuss the fully nonlinear evolution for the first time and examine the crucial influence of SR coupling on the dynamics and the hyperbolicity of the equations of motion (EOMs).
We focus on the coupling function
\begin{equation}
f(\phi)=\frac{\lambda^{2}}{4\kappa}(1-e^{-\kappa\phi^{4}}),\label{eq:coupling}
\end{equation}
where $\lambda,\kappa$ represent coupling parameters.
It satisfies $\frac{d^{2}f}{d\phi^{2}}(0)=0$ so that the BBH does not suffer from linear tachyonic instability.
However, we will show that the BBH becomes unstable against large scalar perturbation and evolves into an SBH.
We have also examined alternative coupling functions, as discussed in \cite{Doneva:2021tvn}, and have observed similar dynamic critical behaviors.

The EOM for gravity can be written as $R_{\mu\nu}-\frac{1}{2}Rg_{\mu\nu}=T^{eff}_{\mu\nu}$
in which the effective energy-momentum tensor
\begin{align}
T^{eff}_{\mu\nu}= & \frac{1}{2}\partial_{\mu}\phi\partial_{\nu}\phi-\frac{1}{4}g_{\mu\nu}(\partial\phi)^{2}\nonumber \\& +2\delta_{\rho\sigma\tau\eta}^{\alpha\beta\gamma\delta}R_{\ \ \gamma\delta}^{\tau\eta}\delta_{\ (\mu}^{\sigma}g_{\nu)\beta}\nabla^{\rho}\nabla_{\alpha}f \nonumber \\
 & -\beta[fG_{\mu\nu}+(g_{\mu\nu}\nabla_{\alpha}\nabla^{\alpha}-\nabla_{\mu}\nabla_{\nu})f]. \label{eq:Tdd}
\end{align}
Here, $\delta_{\rho\sigma\tau\eta}^{\alpha\beta\gamma\delta}$ is the generalized Kronecker delta tensor and $G_{\mu\nu}$ the Einstein tensor.
The scalar field equation reads
\begin{equation}
\nabla_{\mu}\nabla^{\mu}\phi+\frac{1}{2}\left(\beta R+\mathcal{G}\right)\frac{\partial f}{\partial\phi}=0.
\end{equation}

\section{Numerical results}
\subsection{Numerical results of dynamical evolutions}
To study the dynamics of BH in a spherically symmetric spacetime, we use the Painlev\'e--Gullstrand coordinate ansatz
\begin{equation}
ds^{2}=-\alpha^{2}dt^{2}+(dr+\zeta\alpha dt)^{2}+r^{2}(d\theta^{2}+\sin^{2}\theta d\varphi^{2}).\label{eq:PG}
\end{equation}
where $\alpha,\zeta$ represent functions of $(t,r)$, and the apparent horizon $r_{h}$ is determined by $\zeta=1$.
For a Schwarzschild BH, $\alpha=1$, $\zeta=\sqrt{\frac{2M}{r}}$, where $M$ denotes BH mass.
We consider the evolution of a system initially starting with a Schwarzschild BH with mass $M_{0}=1$ and experiencing an initial scalar pulse
\begin{equation}
\phi_{0}=10^{-3}p(r-r_{1})^{2}(r-r_{2})^{2}e^{-\frac{1}{r-r_{1}}-\frac{1}{r_{2}-r}},P_{0}=\partial_{r}\phi_{0},\label{eq:pert}
\end{equation}
when $r\in(r_{1},r_{2})$, and zero otherwise.
We set $r_1=8, r_2=18$.
The parameter $p$ takes control of the perturbation amplitude.
After injection of the initial pulse, the total mass of the system is represented by $M$, and the difference $\delta M=M-M_{0}$ monotonically increases with $p$.
Some other forms of initial pulses have been considered; we qualitatively achieved the same results as those presented in the following sections.
For the numerical method on dynamics, see the appendix.

Without loss of generality, hereafter we set $\lambda=\frac{50}{3}$, $\beta=-\frac{5}{2}$, and $\kappa=1000$.
After the injection of the initial pulse, the resultant spacetime keeps a Schwarzschild BH if $p<p_{s}\approx0.1355610542323$.
When the scalar injection increases its amplitude within the range $p_{s}<p<p_{d}\approx0.2438$, Schwarzschild BHs become unstable and eventually evolve into SBHs.
The terms $p_s$ and $p_d$ represent thresholds for the scalarization and descalarization, respectively, and are computed by the bisection method.
For numerical limitations, improving $p_d$ to higher precision is time-consuming.
When $p>p_{d}$, the infall of strong perturbation makes the BH more massive, and hence the GB term $\mathcal{G}\propto M^{-4}$ becomes considerably small near the horizon.
The induced effective potential barrier cannot confine the scalar hair near the horizon.
Consequently, most of the scalar injection is absorbed by the BH, and finally, the spacetime evolves into a more massive BBH.

\subsection{Dynamical critical behaviors for scalarization}
We fine-tune $p$ around $p_{s}$ and investigate the nonlinear scalarization process to examine the critical dynamics in the scalarization.
Figure~\ref{fig:phip1} shows the scalar field evolution.
Initially, the scalar pulse moves inside and induces violent turbulence near the BBH.
The system then evolves into an intermediate state approximated by the CS at the threshold $p_s$,
\begin{equation}
\phi_{p}(t,r)\approx\phi_{s}(r)+(p-p_{s})e^{\omega_{s}t}\delta\phi_{s}(r)+\text{stable modes},\label{eq:c1}
\end{equation}
where $\phi_{s}(r)$ denotes the scalar distribution of the CS, and $\delta\phi_{s}(r)$ the single unstable eigenmode with positive eigenvalue $\omega_s\simeq 0.106$.
The lifetime of the intermediate state lasts $T_s\propto -\omega_s^{-1}\ln|p-p_s|$.
The final phase of the evolution is determined by the perturbation amplitude.
If $p<p_{s}$, the system evolves into a BBH, but it eventually turns into a stable SBH if $p>p_{s}$.
From Fig.\ref{fig:phip1}, it can be seen that in both cases, gaps exist in the scalar field on the horizon between unstable CS attractor and final stable configurations. The discontinuous change in the scalar field, when $p$ passes through $p_s$, indicates the appearance of a first-order phase transition in the critical scalarization.
The dynamical critical phenomena disclosed here are similar to those in the EMS theory \citep{Zhang:2021nnn,Zhang:2022cmu}.

\subsection{Dynamical critical behaviors for descalarization}
\begin{figure}[h]
\begin{centering}
\includegraphics[width=0.75\linewidth]{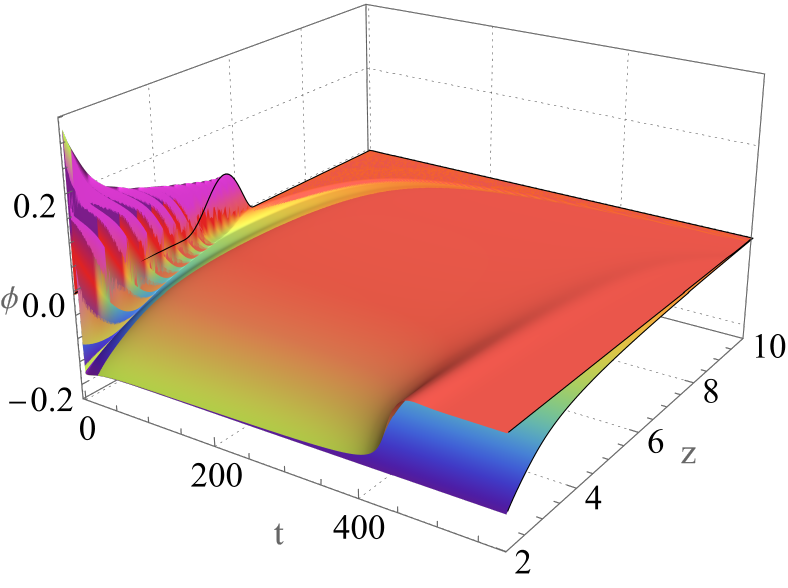}
\par\end{centering}
{\footnotesize{}\caption{{\footnotesize{}\label{fig:phip2}The evolution of $\phi(t,r)$ near the descalarization threshold $p_{d}$. The lower and upper surfaces correspond to the evolution starting with $p=0.2438$ and $p=0.2447$, respectively.}}
}{\footnotesize\par}
\end{figure}

\begin{figure}[h]
\begin{centering}
\includegraphics[width=0.75\linewidth]{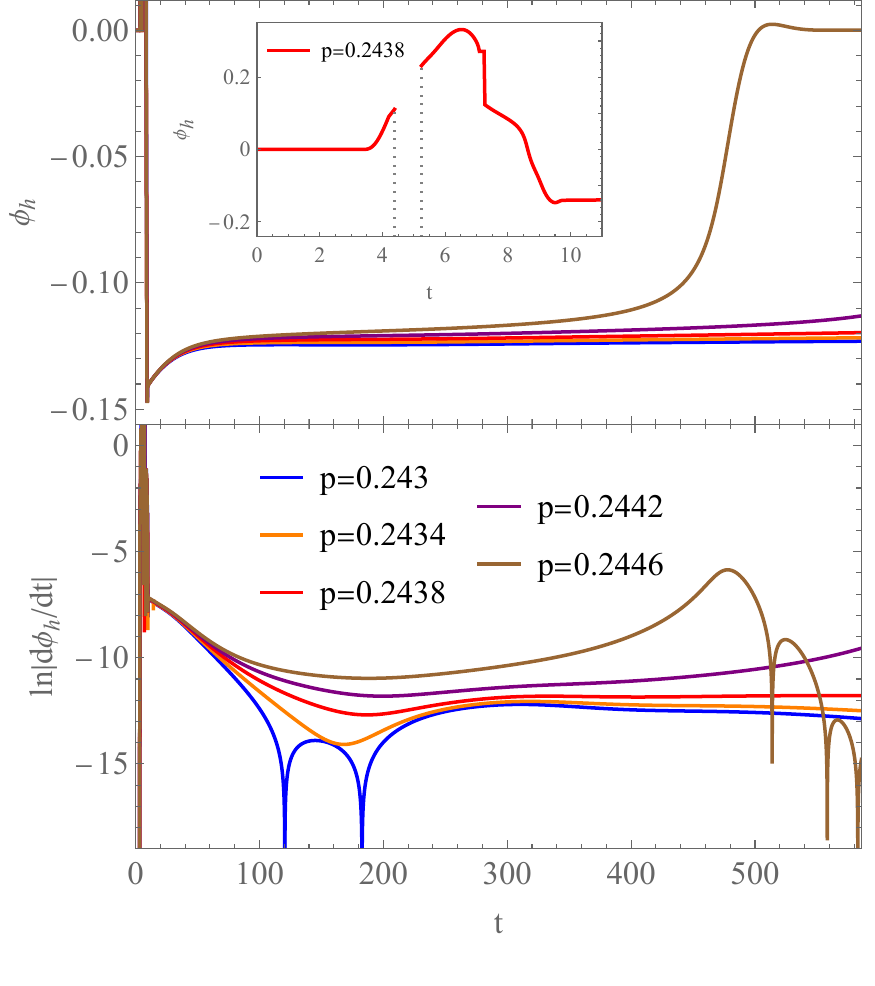}
\par\end{centering}
{\footnotesize{}\caption{{\footnotesize{}\label{fig:phihtp2}The evolution of $\phi_{h}$ and $\ln |\frac{d\phi_h}{dt}|$ near the threshold $p_{d}\approx0.2438$.
A marginal stable attractor exists at the threshold.
The inset shows that for $p=0.2438$, a naked elliptic region appears when $t\in(4.38,5.25)$, and no values of $\phi_{h}$ are obtained during this period.}}
}{\footnotesize\par}
\end{figure}

Now we introduce a strong scalar perturbation to scalarize a BBH, following which we examine the evolution of the SBH when $p$ approaches $p_d$.
Figure~\ref{fig:phip2} shows the evolution of the system.
The system first approaches the CS at $p=p_d$.
Subsequently, if $p<p_d$, the system smoothly and gradually evolves into an SBH.
Figure~\ref{fig:phihtp2} shows the scalar field evolution on the apparent horizon $\phi_h$.
When $p$ approaches $p_d$ from below, $\phi_h$ tends to a constant and evolves extremely slowly.
It is still approximately described in a similar form to that in (\ref{eq:c1}), but with the eigenvalue $\omega_{d}\simeq0$, indicating that the CS near $p_d$ is marginally stable.
However, if $p>p_d$, a jump is observed in the scalar field value, and the scalar hair is observed to be abruptly shedding off.
The SBH finally transforms into a BBH through a first-order phase transition.
\begin{figure}[h]
\begin{centering}
\includegraphics[width=0.75\linewidth]{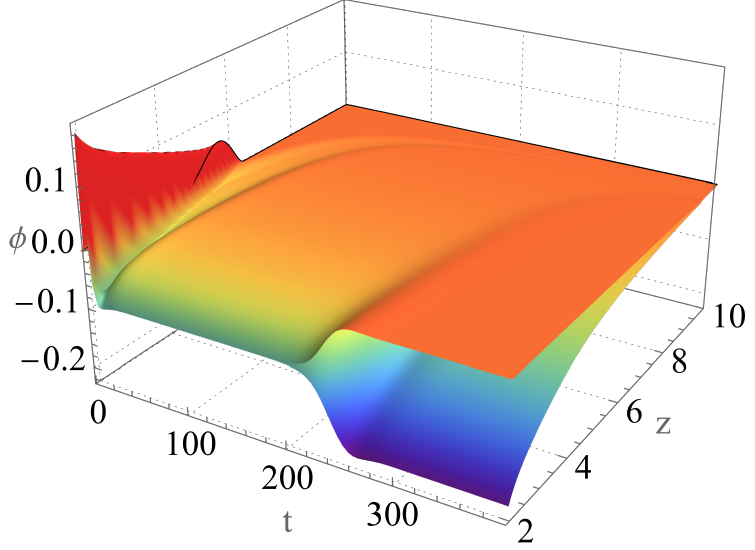}
\par\end{centering}
{\footnotesize{}\caption{{\footnotesize{}\label{fig:phip1}The evolution of $\phi(t,r)$ near the scalarization threshold $p_{s}$. Here, $z=\frac{r}{1+r/L}$ represents the compactified radial coordinate with $L=10$.
The scalar pulse reaches the horizon and first violently interacts with the BBH.
The system then evolves into an attractor.
Finally, the system forms a BBH if $p<p_s$ or an SBH if $p>p_s$.
The upper and lower surfaces correspond to simulations with $p=p_s \mp e^{-28}$, respectively.
}}
}{\footnotesize\par}
\end{figure}

The critical behavior in the descalarization is different from that shown in Fig.\ref{fig:phip1} for the scalarization. It does not agree with that observed in the EMS theory \citep{Zhang:2021nnn,Zhang:2022cmu} either. We observe, for the first time, a marginally stable attractor in descalarization.
This critical dynamical behavior is also distinct from the usual type I critical gravitational collapse, which contains an unstable attractor at the threshold \cite{Choptuik:1996yg,Liebling:1996dx,Bizon:1998kq,Gundlach:2007gc}. It is a new critical dynamical phenomenon disclosed in the first-order phase transition.

\begin{figure}[h]
\begin{centering}
\includegraphics[width=0.75\linewidth]{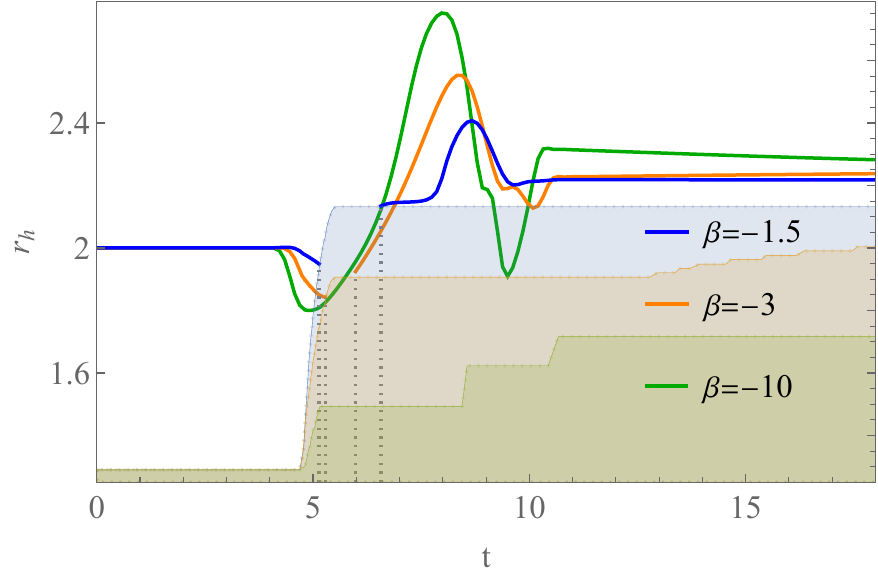}
\par\end{centering}
{\footnotesize{}\caption{{\footnotesize{}\label{fig:exc}The evolution of the apparent horizon $r_h$ and elliptic region for various $\beta$ when $M_0=1,p=0.14$. The light blue, orange, and green regions represent results for $\beta=-1.5,-3,-10$, respectively.}}
}{\footnotesize\par}
\end{figure}

The inset in the upper panel of Fig.\ref{fig:phihtp2} shows the evolution of $\phi_h$ at an early time.
The interval labeled by the dashed lines indicates the presence of a naked elliptic region.
In the sGB theory with other coupling functions, Refs.\citep{Ripley:2019hxt,Ripley:2019irj, Ripley:2020vpk} showed that hyperbolicity often breaks down outside the apparent horizon during evolution.
For hyperbolicity, please see the appendix.
Generally, a naked elliptic region disables numerical simulation.
However, if the apparent horizon hides the elliptic region in time, the pathological region will be prevented from affecting the outside world, and we can continue the simulation.
Figure~\ref{fig:exc} shows the influence of SR coupling on the evolution of BH apparent horizon and elliptic region.
Clearly, an appropriate strength of SR coupling can timely accommodate the apparent horizon to hide the elliptic region and ensure numerical computation.
Nevertheless, if $\beta>-1.5$, the apparent horizon is unable to hide the elliptic region in time, and the code crashes.

\begin{figure}[h]
\begin{centering}
\includegraphics[width=0.7\linewidth]{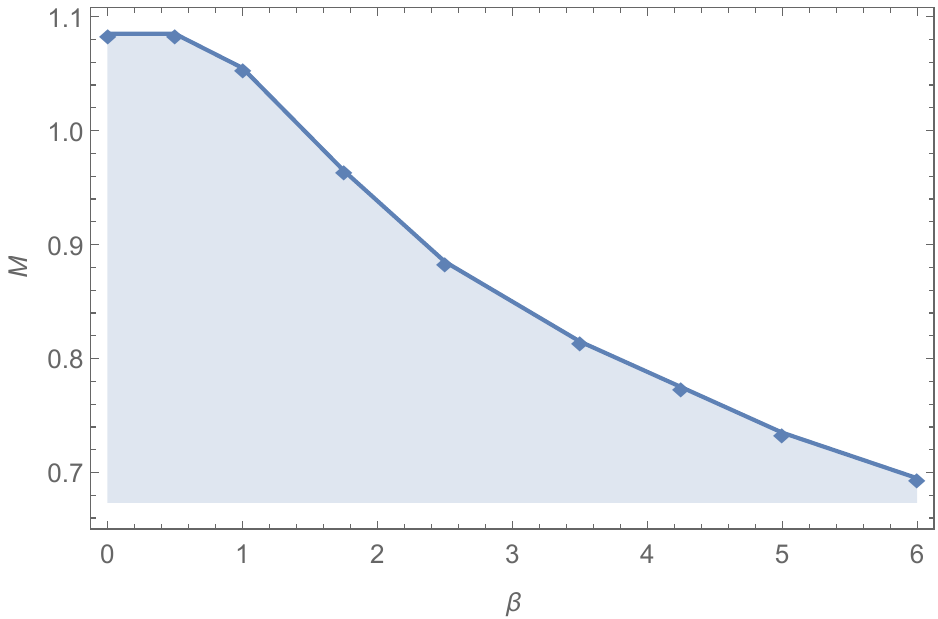}
\par\end{centering}
{\footnotesize{}\caption{{\footnotesize{The parameters space of the existence of a naked elliptic region as a function of $M$ and $\beta$}\label{excision}}}
}{\footnotesize\par}
\end{figure}

We observe that for the given values of $\beta$ and Arnowitt--Deser--Misner (ADM) mass of the spacetime, the elliptic region remains hidden for initial data, i.e., all values of the parameter $p$.
From the shaded area in Fig.\ref{excision}, we can see a naked elliptic region for the selected values of $M$ and $\beta$.
The elliptic region is hidden when the given values of $M$ and $\beta$ are above the line for all initial data. This cannot be obtained in a linear level analysis.

\begin{figure}[ht]
\begin{centering}
\includegraphics[width=0.7\linewidth]{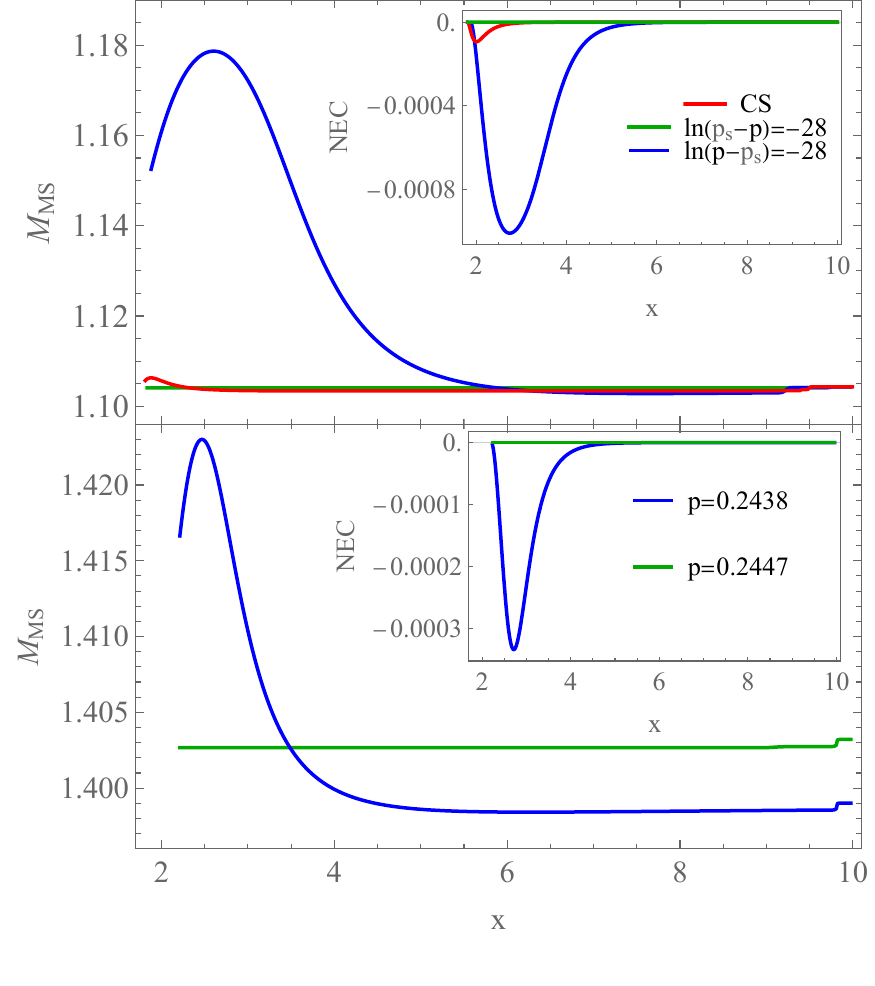}
\par\end{centering}
{\footnotesize{}\caption{{\footnotesize{}\label{fig:p12MSphiX}The distribution of the Misner--Sharp (MS) mass for SBHs (blue), Schwarzschild BHs (green), and
CS (red). The insets indicate the violation of the null energy condition (NEC) for SBHs and CS.}}
}{\footnotesize\par}
\end{figure}

\subsection{Violation of the null energy condition}
Now, we examine the quasilocal MS mass $M_{MS}(t,r)\equiv\frac{r}{2}(1-\nabla_{a}r\nabla^{a}r)=\frac{1}{2}r\zeta^{2}$ \citep{Misner:1964je}.
It equals the radial integral of the effective energy density \citep{Ripley:2019irj}.
For the model we have considered, the effective stress tensor (\ref{eq:Tdd}) does not always satisfy the usual energy conditions such as the NEC $T_{\mu\nu}l^{\mu}l^{\nu}\ge0$ for an outgoing null vector $l^{\mu}=(1,-\alpha,0,0)$. This implies that the MS mass and BH horizon area do not always monotonically increase with radius \citep{Ripley:2019irj,Ripley:2020vpk} and time \citep{Hawking1975,Bardeen1973,Hayward1994,Ashtekar:2004cn}, respectively.
The MS mass distributions in the CS, final SBH, and BBH are shown in Fig.\ref{fig:p12MSphiX}.
The MS mass increases near the horizon and then decreases with radius, implying a negative effective energy density distribution in the space \citep{Ripley:2019irj, Ripley:2020vpk}. Noteworthily, this follows not from the presence of an exotic form of matter but from the synergy of the scalar field coupling with the GB term.

\begin{figure}[h]
\begin{centering}
\includegraphics[width=0.7\linewidth]{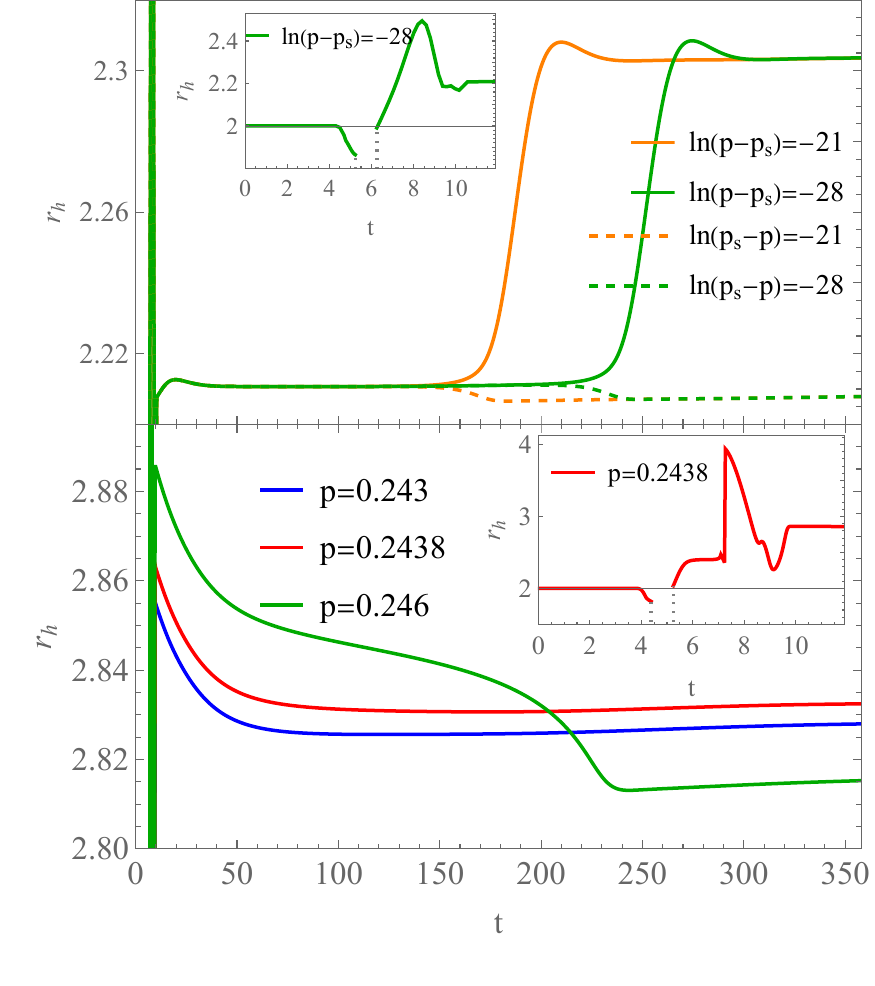}
\par\end{centering}
{\footnotesize{}\caption{{\footnotesize{}\label{fig:p12rht}The evolution of the apparent horizon $r_{h}$ near the thresholds $p_{s}$ and $p_d$.
The upper and lower insets show that naked elliptic regions appear when $t\in(5.23,6.25)$ and $t\in(4.38,5.25)$ and there are no values of $r_{h}$ during these periods for $\ln(p-p_s)=-28$ and $p=0.2438$, respectively.}}
}{\footnotesize\par}
\end{figure}

Figure \ref{fig:p12rht} shows that the BH apparent horizon area can decrease during the evolution. Particularly, the upper panel indicates that the CS has a larger horizon area than the final BBH.
This is not surprising because the violation of NEC causes the breakdown of the BH area theorem \citep{Hawking1975,Bardeen1973,Hayward1994,Ashtekar:2004cn}. Nevertheless, the synergy of a scalar field coupling with both the Ricci scalar and GB invariant affects the final solutions and presents larger Wald entropy than those of the CS and initial BH, as shown in Fig.\ref{fig:MDphihS}.
Besides, we observe that the relation $\ln\dot{r}_{h}\propto2\ln|\dot{\phi}_{h}|$ found in the EMS theory \citep{Zhang:2021nnn,Zhang:2022cmu} does not hold here.

\subsection{Static solutions}
We work out the static solutions and perturbative analysis to more effectively understand the dynamical critical scalarization and descalarization.
For the numerical method, see the appendix.
The green curve in the top panel of Fig.\ref{fig:MDphihS} indicates that the BBH is always a solution.
Interestingly, when the mass is in an appropriate range, there are two extra branches of SBH solutions. The red branch intersects with the BBH at $M=0$ and with the blue branch at point (d) with mass $M_d \simeq 1.396$.
From the perturbative analysis, we learn that the blue and green branches are linearly stable, whereas the red one has an unstable eigenmode with purely positive imaginary frequency shown in the bottom panel.
Noteworthily, the static solution at (d) is marginally stable with zero mode $\omega_I=0$.
The middle panel shows the Wald entropy of the three branches, considering the nonminimal couplings of the scalar field with the Ricci scalar and GB term \citep{Wald:1993nt,Iyer:1994ys}:
\begin{equation}
S=\pi r_{h}^{2}[\beta f(\phi_{h})+1]+4\pi f(\phi_{h}).
\end{equation}
The entropy of each of the three branches increases with $M$.
The differences between them are insignificant and are shown in the inset.
The unstable SBH branch always has the lowest entropy.
It intersects with the stable SBH branch at the cusp (d) of the swallow tail.

\begin{figure}[h]
\begin{centering}
\includegraphics[width=0.75\linewidth]{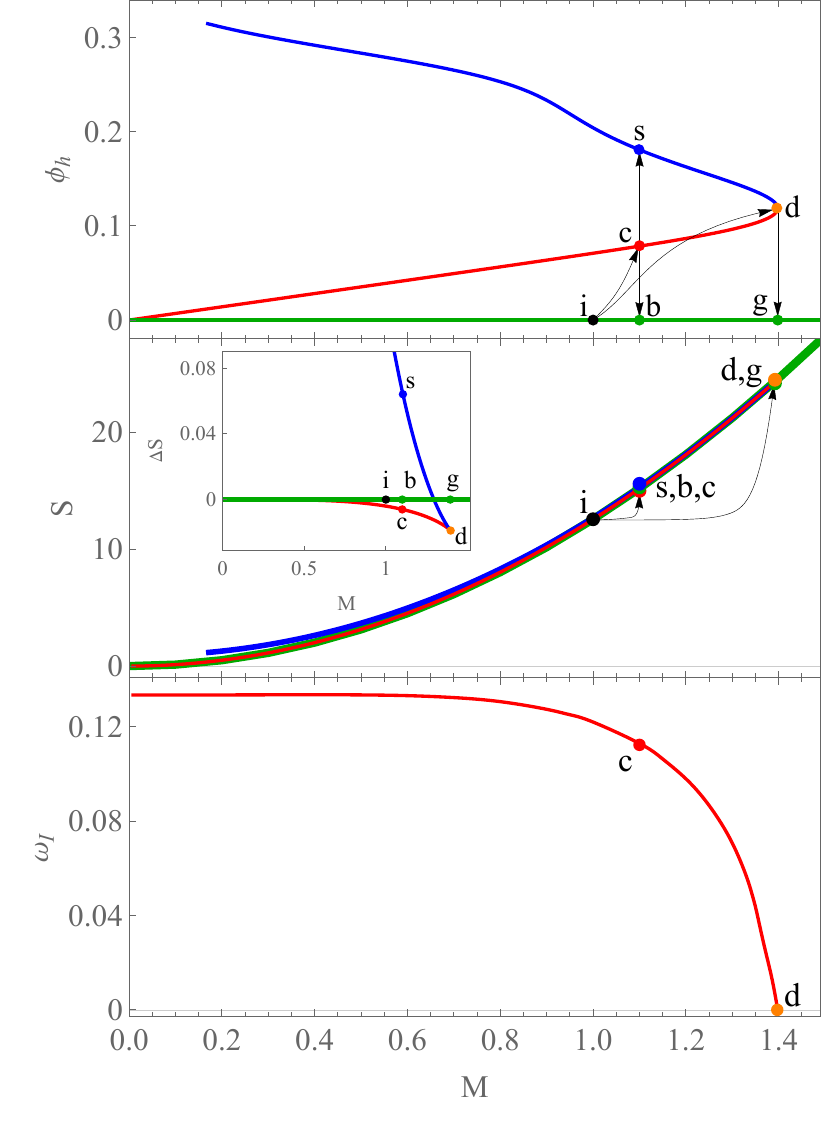}
\par\end{centering}
{\footnotesize{}\caption{{\footnotesize{}\label{fig:MDphihS}The scalar value on the horizon $\phi_{h}$, entropy $S$, and the single unstable eigenmode frequency $\omega_I$ of the static solutions versus ADM mass $M$.
The green, blue, and red lines are the results for BBH, stable SBH, and unstable SBH solutions, respectively.
No stable SBH exists when $M<0.165$.
The inset shows the entropy difference $\Delta S=S-S_B$ in which $S_B=4\pi M^2$ represents the entropy of the BBH.
The upper and lower surfaces in Fig.\ref{fig:phip1} correspond to schematic paths (i,c,b) and (i,c,s), respectively.
The lower and upper surfaces in Fig.\ref{fig:phip2} correspond to schematic paths (i,d) and (i,d,g), respectively. }}
}{\footnotesize\par}
\end{figure}

Now, we compare the dynamic results with the static solutions. The dynamical simulation starts with an initial BBH with $M_0=1$ under perturbation (\ref{eq:pert}) with an amplitude $p$.
When $p$ increases toward $p_s$ from below, the final state stays on the green curve. The BBH becomes more massive in this process.
When $p$ approaches $p_s$, the scalar field starts to adhere to the BBH and transforms the BBH at (i) into an unstable SBH at (c).
If $p<p_s$, the unstable SBH settles down to a heavier BBH described at point (b).
However, if $p>p_s$, the unstable SBH evolves into a stable SBH at (s).
The static solutions at points (c), (b), and (s) have the same mass $M_s=1.098$.
There are gaps between scalar hair values in comparing the unstable SBH at (c) with the BBH at (b) or with the stable SBH at (s). The unstable mode has the eigenvalue $\omega_I\simeq0.110$, in agreement with the $\omega_s$ obtained in the evolution within the error tolerance.
The appearance of the unstable SBH attractor repeats the property disclosed in the EMS theory ~\cite{Zhang:2021nnn}.
When $p$ approaches $p_d$ from below, the SBH becomes marginally stable with zero mode at (d), agreeing with $\omega_d\simeq0$ computed in the dynamical evolution.
Once $p$ passes across $p_d$, the final state jumps to the BBH at point (g), where the scalar hair is suddenly deprived. The BBH at (g) has the same mass as that of the marginally stable SBH at (d).
At this transition point, the marginally stable CS acts as an attractor in the dynamical evolution.

\section{Summary and discussion}
For the class of theories containing a scalar field interacting with the GB invariant, we demonstrated that SR coupling improves the hyperbolic nature in a fully dynamical evolution.
We showed that such Ricci coupling has a positive effect in exploring hair formation and removal in dynamics.
We disclosed the critical dynamics in scalarization and descalarization in a theory with scalar--GB coupling.
We showed that the scalarization of a Schwarzschild BH exhibits a typical type I dynamical critical behavior, in which an unstable attractor lives at the threshold, and a linearly stable BBH can be transformed through this CS and finally settle down as a stable SBH.
In the descalarization, we uncovered a new critical phenomenon, in which the CS is a marginally stable SBH acting as an attractor in the evolution.
This is a novel critical phenomenon in a first-order phase transition. It is distinct from that observed in the EMS theory \citep{Zhang:2021nnn,Zhang:2022cmu}.
Regarding thermodynamics, we observed that dynamical critical descalarization occurs at the cusp of the swallow tail in the entropy--mass relation, where the static unstable SBH branch and the stable SBH branch meet and a marginally stable SBH with zero mode lives.
In the EMS theory, dynamical descalarization occurs before the cusp \citep{Zhang:2022cmu}.
To the best of our knowledge, this is the first example of marginal type I dynamical BH transition.

Considering that an abrupt first-order phase transition in some gravitational systems can induce gravitational wave (GW) emission \cite{Most:2018eaw,Bauswein:2018bma,Weih:2019xvw,Zha:2020gjw}, it is of great interest to further investigate critical dynamics in the first-order phase transition in the generalized theory, whose mechanism could play an important role in GW emission.

\section{Acknowledgments}
This research was supported by the National Key R\&D Program of China under
Grant No. 2020YFC2201400, the Natural Science Foundation of China
under Grant Nos. 11975235, 12005077, and 12035016, and Guangdong Basic
and Applied Basic Research Foundation under Grant No. 2021A1515012374.
B. W. was partially supported by NNSFC under Grant No. 12075202.
Some of our calculations were performed using the tensor-algebra bundle xAct~\href{http://www.xact.es/}{http://www.xact.es/}.

\begin{appendices}
\section{Details for numerical simulation}

We first show the details for full nonlinear dynamical simulation. Defining auxiliary variables
\begin{equation}
Q=\partial_{r}\phi,\ \ \ P=\frac{1}{\alpha}\partial_{t}\phi-\zeta Q.\label{eq:aux}
\end{equation}
we obtain the evolution equations
\begin{align}
E_{\phi}\equiv & \partial_{t}\phi-\alpha\left(P+Q\zeta\right)=0,\label{eq:phit}\\
E_{Q}\equiv & \partial_{t}Q-\partial_{r}\left[\alpha(P+Q\zeta)\right]=0,\label{eq:Qt}\\
E_{P}\equiv & \partial_{t}P-A_{P}=0,\label{eq:Pt}\\
E_{\zeta}\equiv & \partial_{t}\zeta-A_{\zeta}=0,\label{eq:zetat}
\end{align}
and the constraint equations
\begin{equation}
\alpha'=C_{\alpha},\zeta'=C_{\zeta}.\label{eq:ctr}
\end{equation}
Here $A_{P,\zeta},C_{\alpha,\zeta}$ are lengthy expressions of $\phi,P,Q,\alpha,\zeta$
and their radial derivatives.
Given initial data $\phi_{0},P_{0}$, we can work out the initial $Q$ from (\ref{eq:aux}) and $\alpha,\zeta$ from (\ref{eq:ctr}). Then by using (\ref{eq:phit}, \ref{eq:Qt}, \ref{eq:Pt}), we can work out the $\phi,Q,P$ on the next time slice. Iterating this procedure, we obtain all variables on all time slices. The equation (\ref{eq:zetat}) is auxiliary. The constraints (\ref{eq:ctr}) are solved by the Newton--Raphson method. The evolution equations (\ref{eq:phit}, \ref{eq:Qt}, \ref{eq:Pt}) are solved by the method of lines, in which the radial derivative is discretized with fourth-order finite difference method, and the time direction is evolved with fourth-order Runge--Kutta method.
The Courant--Friedrichs--Lewy condition is satisfied. Kress--Oliger dissipation is employed to stabilize the code.
The numerical implementation, written in Mathematica, is available in the GitHub repository~\href{https://github.com/hours1127/hairy-black-hole-descalarization.}{https://github.com/hours1127/hairy-black-hole-descalarization.}.

With the gauge freedom $\alpha\to l\alpha,t\to t/l$ in the metric, we can set $\alpha|_{r\to\infty}=1$. This implies that we take the time coordinate $t$ as the proper time of the observer at spatial infinity. The metric function $\zeta\to\sqrt{\frac{2M}{r}}$ at a large radius, where $M$ represents the total mass of the spacetime.
To improve the stability of our numerical code, we use the variable $s=\sqrt{r}\zeta$ and impose the boundary condition $s|_{r\to\infty}=\sqrt{2M}$ in the code.

We compactify the radial direction as $z=\frac{r}{1+r/L}$, where $L=10M_{0}$, and discretize $z$ with a uniform grid. On each time slice, the computational domain covers $z\in[z_{e},L]$, where $z_{e}$ denotes the inner boundary and is located at some distance inside the apparent horizon.

\begin{figure}[h]
\begin{centering}
\includegraphics[width=0.9\linewidth]{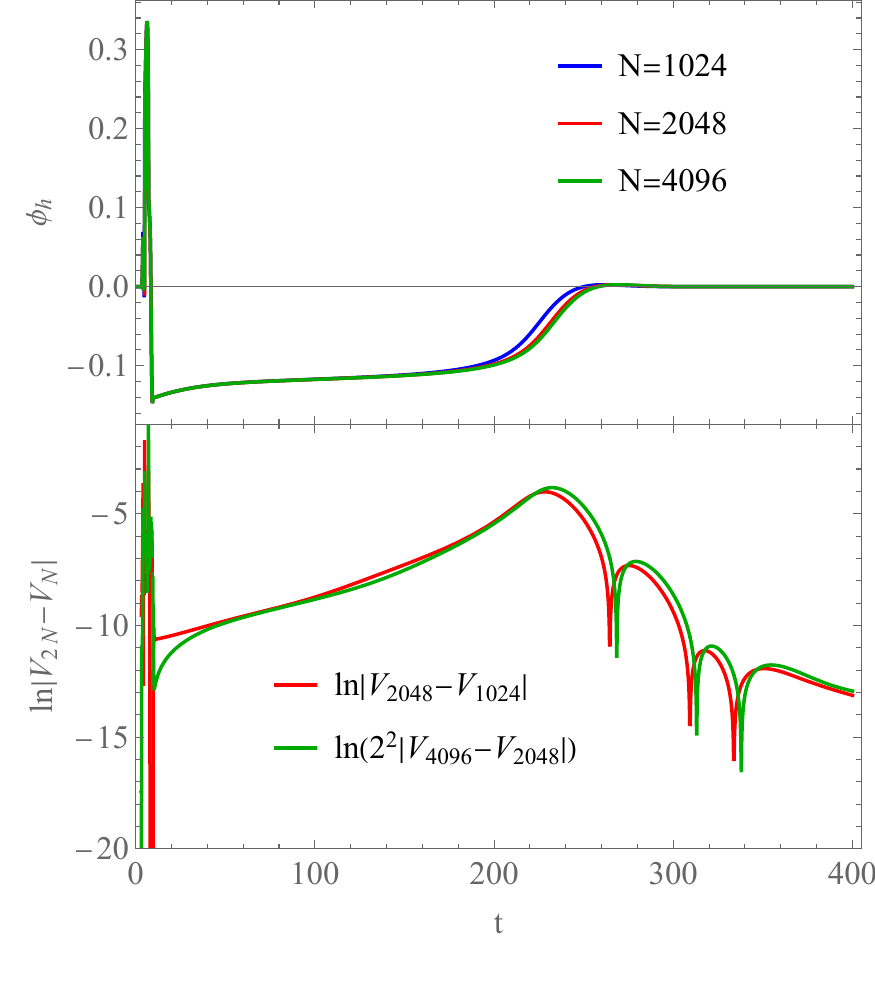}
\par\end{centering}
{\footnotesize{}\caption{{\footnotesize{}\label{fig:converge}The evolution of $\phi_h$ with different grid points $N$ when $p=0.246$. The differences $V_{2048} - V_{1024}$ almost overlaps with the four times of $V_{4096} - V_{2048}$.}}
}{\footnotesize\par}
\end{figure}

The convergence order $q$ of our method can be estimated by $\frac{V_{2N}-V_{N}}{V_{4N}-V_{2N}}=2^{q}+O(1/N)$, where $V_{N}$ represents the result obtained with $N$ radial grid points. As shown in Fig.\ref{fig:converge}, our method converges to the second order. The dynamical simulations shown in the main text are obtained with $N=4096$.

\section{Hyperbolicity}
Now, let us consider the hyperbolicity of the scalar field equation.
The radial characteristic speeds $c_{\pm}=\mp\xi_{t}/\xi_{r}$ of the scalar field $\phi$ are determined by the characteristic equation
\begin{equation}
\det\left[\left(\begin{array}{cc}
\frac{\delta E_{P}}{\delta(\partial_{a}P)} & \frac{\delta E_{P}}{\delta(\partial_{a}Q)}\\
\frac{\delta E_{Q}}{\delta(\partial_{a}P)} & \frac{\delta E_{Q}}{\delta(\partial_{a}Q)}
\end{array}\right)\xi_{a}\right]=0,
\end{equation}
where $\xi_{a}=(\xi_{t},\xi_{r})$. This equation has the quadratic
form $\mathcal{A}c^{2}+\mathcal{B}c+\mathcal{C}=0$, where $\mathcal{A},\mathcal{B},\mathcal{C}$
are complicated functions of the metric and scalar field \citep{Ripley:2019hxt,Ripley:2019irj, Ripley:2020vpk}. The regions of the spacetime where the discriminant $\mathcal{D}\equiv\mathcal{B}^{2}-4\mathcal{AC}<0$ are elliptic regions.
If an elliptic region appears, we excise it and shift the inner boundary $z_{e}$
at the next time slice.

\section{Details for solving static solutions and perturbations}

Last, we present the numerical procedure for solving the static solutions and quasinormal frequencies.
For convenience, we use the following metric ansatz \cite{Blazquez-Salcedo:2018jnn,Silva:2018qhn}:
\begin{eqnarray}
ds^2&=&-e^{\Lambda(r)+\epsilon F_t(t_s,r)}dt_s^2+e^{T(r)+\epsilon F_r(t_s,r)} dr^2\nonumber\\&&+r^2(dr^2+\sin^2\theta d\varphi^2),\label{eq:ansatz-2}\\
\phi(t_s,r)&=&\phi_0(r)+\epsilon \phi_1(t_s,r).
\end{eqnarray}
were $\epsilon$ represents an infinitesimal perturbation parameter.
The zeroth-order terms in the EOMs govern the static background functions $\Lambda$, $T$, and $\phi_0$, while the first-order terms govern the time-dependent perturbation functions $F_t,~F_r$ and $\phi_1$.
For a static case, the coordinate transformation between Painlev\'e--Gullstrand coordinate ansatz (Eq.(5)) and~(\ref{eq:ansatz-2}) takes the form $dt_s=dt-\frac{\zeta dr}{(1-\zeta^2)\alpha}$.
Noteworthily, $\phi_h,S$ and $\omega_I$ do not change under this transformation.

From the zeroth-order terms, the field equations of $\Lambda$, $T$, and $\phi_0$ are cast into three coupled ordinary differential equations (ODEs).
After some algebra calculations, we obtain two coupled second-order ODEs for $\Lambda$ and $\phi_0$ and an algebraic equation for $T$.
We compactify the radial coordinate as $x=1-\frac{r_h}{r}$.
The boundary conditions are imposed as $e^{\Lambda}|_{x\rightarrow 0}\rightarrow 0$ and $e^{-T}|_{x\rightarrow 0}\rightarrow 0$ on the BH horizon and as $\phi_0|_{x\rightarrow 1}\rightarrow 0$ at infinity.
After solving the equations, one can rescale $e^{\Lambda}|_{x\rightarrow 1}\rightarrow 1$ at infinity by gauge freedom.
The asymptotic behavior of the scalar field takes the form $\phi\sim\frac{D}{r}+O(r^{-2})$, where $D$ denotes the scalar charge.
Specifying $D,\lambda,\kappa,\beta$ and $r_h$, we can obtain $\Lambda$ and $\phi_0$ by the shooting method.
The metric function $T$ is then obtained as a function of $\Lambda$ and $\phi_0$.

From the first-order terms, after algebraic calculation, the scalar perturbation $\phi_1$ decouples from the metric perturbations $F_t$ and $F_r$ \cite{Blazquez-Salcedo:2018jnn,Silva:2018qhn}. It has a single second-order differential equation, which can be used to compute the quasinormal modes of the background solutions.
With the decomposition $\phi_1(t_s,r)=\psi_1(r) e^{i\omega t_s}$, the scalar perturbation equation reads
\begin{equation}
[\partial_r^2+A(r)\partial_r +(\omega^2B(r)^2-U(r))]\psi_1(r)=0.
\end{equation}
where the coefficients $A(r),B(r)$, and $U(r)$ represent complicated functions depending on the background metric and scalar field $\Lambda$, $T$, and $\phi_0$.
We will focus on perturbations with purely imaginary eigenfrequencies that $\omega=i \omega_I$.
$\omega_I$ can be obtained with the same method as that described in \cite{Blazquez-Salcedo:2018jnn,Silva:2018qhn}.

\end{appendices}

 \end{document}